\documentclass[11pt]{article}
\input{amssym.def}
\input{amssym}
\newtheorem{theorem}{Theorem}
\newtheorem{proposition}{Proposition}
\newtheorem{lemma}{Lemma}

\newcommand{\begsection}[1]{\setcounter{equation}{0}\section{#1}}

\def\R{{\mathcal R}}

\def\quadratino{\hfill
\vbox{\hrule\hbox{\vrule\vbox to 7 pt {\vfill\hbox to 7 pt
{\hfill\hfill}
\vfill}\vrule}\hrule}\par}
\def\R{\Bbb R} 
\def\Z{\Bbb Z} 
\def\N{\Bbb N}
\def\T{\Bbb T}  
\def\C{\Bbb C}

\def\Sc{Schr\"o\-din\-ger}
\def\hp{{\hbar}}

\def\la{\langle}
\def\be{\begin{equation}}
\def\ee{\end{equation}}
\def\ra{\rangle}
\def\ds{\displaystyle}

\def\om{\omega}
\def\Om{\Omega}
\def\ep{\epsilon}

\def\As{{\mathcal A}_{\om,\sigma}}

\begin{document}
\baselineskip=19pt
\begin{center}
{\large\bf A UNIFORM  QUANTUM VERSION OF THE \\
CHERRY THEOREM}
\end{center}
\vskip 13pt
\begin{center}
  Sandro Graffi\footnote{Dipartimento di Matematica, 
Universit\`a di Bologna
(graffi@dm.unibo.it)},  
Carlos Villegas Blas\footnote{
Instituto de  Matematica,  Universit{a}d Nacional de Mexico,  Cuernavaca (Mexico).}  
\end{center}
\begin{abstract}
\noindent
Consider in $L^2(\R^2)$ the  operator family
$H(\epsilon):=P_0(\hbar,\omega)+\epsilon F_0$.  
$P_0$ is the quantum harmonic oscillator with diophantine frequency vector
$\om$, $F_0$ a  bounded 
pseudodifferential operator with symbol  
decreasing to  zero at infinity in phase space, and $\ep\in\C$.  Then there exist 
$\ep^\ast >0$ independent of $\hbar$ and an open set
$\Omega\subset\C^2\setminus\R^2$  such that if
$|\ep|<\ep^\ast$ and $\om\in\Om$ the quantum normal form near $P_0$
converges uniformly with respect to $\hbar$. This yields an exact
quantization formula for the eigenvalues, and for $\hbar=0$ the classical
Cherry theorem on convergence of Birkhoff's normal form for complex
frequencies is recovered.
\end{abstract}
\vskip 1cm   
%
 
%


\begsection{Introduction and statement of the results}
\setcounter{equation}{0}%
\setcounter{theorem}{0}%
\setcounter{proposition}{0}%
\setcounter{lemma}{0}%
\setcounter{corollary}{0}%
\setcounter{definition}{0}%
Consider  in the phase space $\R^{2l}$ with canonical coordinates denoted 
$(x,\xi)$ the Hamiltonian system defined by the principal
function
\begin{eqnarray}
\label{Ham1}
p_\ep(x,\xi;\om)&:=&p_0(x,\xi)+\ep f_0(x,\xi)
\\
\label{azioni}
p_0(x,\xi;\om)&:=&\frac12(|\xi|^2+|\om x|^2)\,=\,
\sum_{k=1}^l\om_kI_k(x,\xi), 
\\
I_k(x,\xi)&:=&\frac{1}{2\om_k}[\xi_k^2+\om^2_kx_k^2], \quad k=1,\ldots,l.
\end{eqnarray}
Here $f_0:\R^{2l}\to\R$ is analytic; $f_0=O([|\xi|^2+|\om x|^2]^{s/2})$, $s\geq 3$, as $|x|+|\xi|\to 0$, and 
$\ep\in\R$.  Any analytic Hamiltonian near a
non-degenerate elliptic equilibrium point can  be written in the form (\ref{Ham1}).
Let the {\it frequencies} $\om:=(\om_1,\ldots,\om_l)$ fulfill a diophantine
condition, i.e
\begin{equation}
\label{Diofanto}
\la{ \om}, { k}\ra \geq {\gamma}{|{k}
|^{-\tau}}, \quad \forall {
k}\in 
\Z^l\setminus\{0\}, \; |k|:=|k_1|+\ldots+|k_l|,\;
,\;\gamma>0,\;\tau >l-1.
\end{equation}
Under these
circumstances the Birkhoff theorem holds, namely (see e.g.\cite{SM}, \S 30):
\par\noindent
{\it $\forall\,N\in\N$,  $\forall\,p\in\N$, $\forall\,\ep\in\R$ one can construct an analytic, canonical bijection
$(y,\eta)=\chi_{\ep,N}(x,\xi): \R^{2l}\leftrightarrow  \R^{2l}$  and a sequence of analytic
functions $Y_p(I;\om): \R_+^l\to\R$  such that:
\begin{eqnarray}
p_\ep\circ \chi_{\ep,N}^{-1}(y,\eta)=\sum_{k=1}^l\om_kI_k(y,\eta)+\sum_{p=1}^{N-1}
Y_p(I(y,\eta);\om)\ep^p+\ep^{N}R_N(y,\eta;\ep).
\end{eqnarray}
}
The $l$ functions $I:=(I_k(y,\eta): k=1,\ldots,l)$, the mechanical actions, are thus
 first integrals of the transformed Hamiltonian up to an error of order $\ep^N$.
Hence the system is integrable if the remainder in (1.4) vanishes as
$N\to\infty$, namely if the {\it Birkhoff normal form}
\be
\label{Bi}
B(I;\om,\ep):=\la\om,I\ra+\sum_{p=1}^{\infty}Y_p(I;\om)\ep^p, \quad
\la\om,I\ra:=\sum_{k=1}^l\om_kI_k 
\ee
 converges when the actions belong to some ball $|I|<R$ of
$\R_+^l$. However, as proved by C.L.Siegel\cite{Si} in 1941, (\ref{Bi})  is
generically divergent (a particular convergence criterion has been later isolated by
R\"ussmann\cite{Ru}; see also \cite{Ga}. It states that (\ref{Bi}) converges if $Y_p(I,\om)=Y_p(\la\om,I\ra)$).  Already in 1928, on the other hand, T.M.Cherry\cite{Ch} (see 
also \cite{SM}, \S 30; a more recent proof can be found in \cite{Ot}) remarked that, when
$l=2$, the  normal form is convergent provided the frequencies $\omega$ are complex with non
vanishing imaginary part. 
Under this assumption  the small
denominator mechanism which generates the divergence becomes instead a large denominator
one entailing the convergence.

We prove here that  
under the same assumptions on the frequencies, but much more restrictive
conditions on the perturbation, the Cherry theorem holds in quantum mechanics as well, with estimates
uniform with respect to the Planck constant
$\hbar$.  Namely, the quantum Birhoff normal form (see \cite{Sj}) converges uniformly with respect to $\hbar$, and this yields an {\it exact} quantization formula for the quantum spectrum.
\newline
Consider indeed in $L^2(\R^2)$ the
operator
$H(\ep)=P_0(\hbar,\om)+\ep F_0$ under the assumptions:
\begin{itemize}
\item[(A1)]  $P_0(\hbar,\om)$ is the harmonic-oscillator \Sc\ operator  
 with frequencies  $\om$:
\be
\label{HO}
P_0(\hbar,\om)\psi=-\frac12\hbar^2\Delta \psi+\frac12[\om_1^2x_1^2+\om_2^2x_2^2]\psi,
\;\; D(P_0)=H^2(\R^2)\cap L_2^2(\R^2).
\ee
\item[(A2)]  Let $\om_1=a+ib$, $\om_2=c+id$, $a\neq 0, c\neq 0$, $\la\om_1,\om_2\ra:=ac+bd$.
Then $\om\in\Gamma\subset\C^2$, where:
 \be
 \label{Gamma}
\Gamma:=\left\{\om\in\C^2\,|\,0<\delta_1\leq|\om|\leq \delta_2\,|\,\frac{|\la\om_1,\om_2\ra|}{|\om_1\om_2|}=\frac{|ac+bd|}{\sqrt{(a^2+b^2)
(c^2+d^2)}}\leq \delta<1\right\}
\ee
\end{itemize}
To state the assumption on the perturbation $F_0$,   define an analytic action $\Psi:\T^2\times\R^{2}\times\R^2\to\C^{2}\times\C^2,\quad 
(x,\xi) \mapsto  (x^{\prime},\xi^{\prime})=\Psi_{\phi,\om}(x,\xi)$ of $\T^2$ into
$\C^{2}\times \C^2$ through the flow of $p_0(\cdot,\om)$ of {\it real} initial data $u:=(x,\xi)\in\R^{2}\times\R^2$ but {\it complex} frequencies $\om\in\Gamma$:
\begin{eqnarray}
\label{aazione}
\left\{\begin{array}{l}
{\ds x^{\prime}_k:=\frac{\xi_k}{\om_k}\sin\phi_k+x_k\cos\phi_k,}
\\
{\;\xi^{\prime}_k:=\xi_k\cos\phi_k-\om_kx_k\sin\phi_k,}
\end{array}\right. \qquad k=1,2
\end{eqnarray}
Let $f(z)\in C(\C^{2}\times\C^2;\C)$.  Then $f\circ \Psi_{\phi,\om}(x,\xi)=f\circ\Psi_{\phi,\om}(u)$, denoted $f_{\phi,\om}(u)$,   is a $\phi-2\pi$- periodic function $\forall\,(u,\om)\in\R^2\times\R^2\times\Gamma$ fixed.  We further denote: $f(u):=f_{\phi,\om}|_{\phi=0}$ and
\vskip 3pt\noindent
1.  $f_{\nu,\om} (u)$ the Fourier coefficients of $ f_{\phi,\om}(u)$ :
$$
f_{\nu,\om} (u):=\frac{1}{(2\pi)^{{2}}}\int_{\T^2}f(\Psi_{\phi,\om}(u))e^{-i\la
\nu,\phi\ra}\,d\phi, \quad \nu\in\Z^2.
$$
2.
\be
\label{FC}
\widehat{f}_{\nu,\om}(s) := \frac{1}{2\pi} \int_{\R^{2}\times\R^2} 
 f_{\nu,\om}(u)e^{-i\la s,u\ra}\, du \,.
\ee
their space Fourier transform. Here $\widehat{g}(s)$ is the Fourier transform of $g$:
 $$
 \widehat{g}(s)=\frac1{2\pi}\int_{\R^{2}\times\R^2}g(u)e^{-i\la s,u\ra}\,du\,,\quad g(u)\in L^1(\R^{2}\times\R^2).
 $$
\par\noindent
3.  ${\cal F}_{\sigma}:=\{f\in L^1(\R^2\times \R^2)\,|\,\|f\|_\sigma <+\infty\}$,  $\sigma >0$.  Here:
\be
\label{normas}
\|f\|_{\sigma}:=\int_{R^{2}\times\R^2}|\widehat{f}(s)|e^{\sigma |s|}\,ds<+\infty
\ee 
\par\noindent
4.  ${\cal A}_{\Gamma,\rho,\sigma}:=\{f\in L^1(\R^2\times \R^2)\cap C(\C^2\times \C^2)\,|\,\|f\|_{\Gamma,\rho,\sigma}<+\infty\}$,  $\rho>0$, $\sigma >0$. Here:
\begin{eqnarray}
\label{norma4}
\|f\|_{\Gamma,\rho,\sigma}:=\sup_{\om\in\Gamma}\sum_{\nu\in\Z^2}\,e^{\rho|\nu|}\|f_{\nu,\om}\|_{\sigma};
\end{eqnarray}
We can now state our assumption on the perturbation.
\begin{itemize}
\item[(A3)]   $F_0$ is a semiclassical pseudodifferential operator of order $\leq 0$ with  (Weyl) symbol 
$f_0\in{\cal A}_{\Gamma,\rho,\sigma}$  for some $\rho>0$, $ \sigma>0$. Explicitly:
(notation as in \cite{Ro}) $F_0= Op^W_h(f_0)$, 
\begin{eqnarray}
\label{Weyl}
(F_0\psi)(x)=\frac{1}{h^2}\int\!\!\!\int_{\R^2\times\R^2}e^{i\la
(x-y),\xi\ra/\hbar} f_0((x+y)/2,\xi)\psi(y)\,dyd\xi,\quad \psi\in{\cal S}(\R^2).
\end{eqnarray}
 \end{itemize}
 {\bf Remarks}
\begin{enumerate} 
\item
Since (\cite{Ro}, \S II.4) $ \|F\|_{L^2\to L^2}\leq \|\widehat{f}\|_{L^1}$, 
  $F_0$  extends to a conti\-nuous 
operator in $L^2(\R^2)$ because:
\be
\label{cont}
\|F_0\|_{L^2\to L^2} \leq 
\|\widehat{f}_0\|_{L^1} \leq \|f_0\|_{\sigma}\leq \|f_0\|_{\Gamma,\rho,\sigma}.
\ee
\item 
Any $f\in {\cal A}_{\Gamma,\rho,\sigma}$ admits a holomorphic continuation from $u=(x,\xi)\in\R^2\times \R^2$  to the strip $\{z=(z_1,z_2)\in\C^2\times\C^2\,|\, |{\rm Im}\,z|<\sigma\} $. Obviously this holomorphic continuation can be different from the function $f\circ\Psi_{\phi,\om}(z_1,z_2):\C^2\times \C^2\to\R$, as in the example  $f=e^{-|z|^2}P(z):\C^2\times \C^2\to\C$, $P$ any polynomial, discussed in Appendix.  
\end{enumerate}
Since $F_0$ is bounded, $H(\ep)$  defined on $D(P_0)$ is closed with
 pure-point  spectrum $\forall\,\ep\in\C$, and is self-adjoint for $\ep\in\R$ if 
$\om\in\R_+^2$.  Moreover, 
$P_0$ can be considered a  semiclassical pseudodifferential operator of order $2$ with
symbol $p_0(x,\xi;\om)$.
\vskip 0.3cm\noindent
\begin{theorem}
\label{mainth}
Let (A1-A3) be verified and let $h^\ast>0$.  Then 
 there exists $\ep^\ast >0$ 
independent of $\hbar\in[0,\hbar^\ast]$
 such  that if $|\ep|<\ep^\ast$ the spectrum of 
$H(\ep)$  is given by the quantization formula
\begin{eqnarray}
\label{quantiz}
E_n(\hbar,\ep)&=&\la
\om,n\ra\hbar+\frac12(\om_1+\om_2)\hbar+
{\cal  
N}(n\hbar,\hbar;\ep).
\\
\label{nserie}
{\cal  N}(n\hbar,\hbar;\ep)&=&\sum_{p=1}^\infty \,{\cal 
N}_p(n\hbar,\hbar)\ep^p
\end{eqnarray}
Here $n=(n_1,n_2) $, $n_i=0,1,\ldots$, and:
\par\noindent
1. ${\cal  N}_p(I,\hbar):\R^2_+\times[0,h^\ast]\to\C$ is 
analytic in 
$I$ and continuous in $\hbar$;
\par\noindent
2.  The series (\ref{quantiz}) has convergence radius $\ep^\ast$ uniformy with respect to
$(I,\hbar)\in\Om\times [0,h^\ast]$. Here $\Om$ is any compact of $\R^2_+$;
\par\noindent
3. ${\cal  N}_p(I,\hbar): p=1,2,\ldots$ admits an asymptotic expansion to all orders in
$\hbar$; the order $0$ term is the coefficient $Y_p(I)$ of the Birkhoff normal form. 
\end{theorem}
\noindent
{\bf Remarks}
\begin{enumerate}
\item The conditions of the Cherry theorem   are much less
restrictive than the present ones. In particular, the standard Schr\"odinger
operator in which $f_0$ depends only on $x$ is excluded. On the other hand, in the classical case
$\hbar=0$ we obtain an improved version of the theorem: indeed, in our conditions  the Birkhoff normal form converges, for $\epsilon$ small enough,
in {\it any} compact of $\R^2$. To our knowledge this result is new.
\item  Taking  $\hbar=0$ in ${\cal  N}_p(I,\hbar)$ (\ref{quantiz}) becomes $\ds
E_\nu^{BS}(\hbar,\ep):=\la\om,n\ra\hbar+\frac12(\om_1+\om_2)\hbar+\sum_{p=1}^\infty
\,Y_p(n\hbar)\ep^p$, namely the Bohr-Sommerfeld quantization of the Birkhoff normal form.
Formula (\ref{quantiz}) yields all corrections needed to recover the eigenvalues
$E_n(\hbar,\ep)$. 
\item For any fixed $n$ and $\hbar$ the series (\ref{quantiz}) coincides with the  Rayleigh-Schr\"odinger perturbation expansion near the simple eigenvalue $\ds \la\om,n\ra\hbar+\frac12(\om_1+\om_2)\hbar$ of $P_0$ \cite{GP}.
\item The eigenvalues $E_n(\hbar,\ep)$ admit the interpretation of quantum 
resonances of a self-adjoint Schr\"odinger operator. For this matter the reader is referred to
\cite{MS}, where under much more general conditions on $f_0$ the eigenvalues are obtained by
an exact quantization of the KAM iteration scheme.
 \end{enumerate}
 We thank  Dario Bambusi for a critical reading of the manuscript and Andr\'e Martinez for providing us a first proof of Lemma 2.3. 
 \begsection{Proof of the results}
\setcounter{equation}{0}%
\setcounter{theorem}{0}%
\setcounter{proposition}{0}%
\setcounter{lemma}{0}%
\setcounter{corollary}{0}%
\setcounter{definition}{0}%
The proof is to be obtained in four steps.
\par\noindent {\it 1. Perturbation theory: the formal construction}
\par\noindent
Look for a unitary 
transformation $\ds U(\om,\ep,\hbar)=e^{i W(\ep)/\hbar}:
L^2\leftrightarrow L^2$,
$W(\ep)=W^\ast(\ep)$, $\ep\in\R$, such that: 
\begin{eqnarray}
\label{passo1bis}
S(\ep):=UH(\ep)U^{-1}=P_0(\hbar,\om)+\ep Z_1+\ep^2 Z_2+\ldots+
\ep^k R_k(\ep)
\end{eqnarray}
where $[Z_p,P_0]=0$, $p=1,\ldots,k-1$. Recall the formal commutator expansion: 
\be
\label{commu}
e^{it W(\ep)/\hbar}He^{-it W(\ep)/\hbar}=\sum_{l=0}^\infty t^lH_l,\quad H_0:=H,\quad
H_l:=\frac{[W,H_{l-1}]}{i\hbar l}, \;l\geq 1
\ee
Looking for $W(\ep)$ under the form of a power series, 
$W(\ep)=\ep W_1+\ep^2W_2+\ldots$,   (\ref{commu}) becomes:
\be
\label{Explicitq1}
S=\sum_{s=0}^{k}\ep^s  P_s +\ep^{k+1}{R}^{(k+1)}
\ee
where
\be
\label{Explicitq2}
{P}_s=\frac{[W_s,P_0]}{i\hbar}+F_s,\quad s\geq 1, \;F_1\equiv F_0
\ee
\begin{eqnarray*}
F_s &=&\sum_{r=2}^s\frac{1}{r!}\sum_{{j_1+\ldots+j_r=s}\atop {j_l\geq
1}}\frac{[W_{j_1},[W_{j_2},\ldots,[W_{j_r},P_0]\ldots]}{(i\hbar)^r} 
\\
&+&\sum_{r=2}^{s-1}\frac{1}{r!}\sum_{{j_1+\ldots+j_r=s-1}\atop {j_l\geq
1}}\frac{[W_{j_1},[W_{j_2},\ldots,[W_{j_r},F_0]\ldots]}{(i\hbar)^r} 
\end{eqnarray*}
Since  $F_s$ depends on $W_1,\ldots,W_{s-1}$,  (\ref{passo1bis}) yields the
recursive homological equations:
\be
\label{qhomeq}
\frac{[W_s,P_0]}{i\hbar} +F_s=Z_s, \qquad [P_0,Z_s]=0
\ee
To solve for $S$, $W_s$, $Z_s$, we can equivalently look for their symbols; from now on, 
we denote by the same letter, but in small case, the symbol $\sigma(A)$ of an operator $A$, except for the symbol of $S$, denoted $\Sigma$. Let us now recall the following relevant results (see e.g.
\cite{Fo},\S $3.4$): 
\begin{enumerate}
\item $\ds \sigma([A,B]/i\hbar)=\{a,b\}_M$, where $ \{a,b\}_M$ is the Moyal bracket of $a$ and
$b$. 
\item
Given $(g,g^{\prime})\in\As$, their Moyal
bracket $\{g,g^{\prime}\}_M$ is defined as
$$
\{g,g^{\prime}\}_M=g\# g^{\prime}-g^{\prime}\#g,
$$
where $\#$ is the composition of  $g, g^{\prime}$ considered as Weyl
symbols.
\item In the Fourier transform representation, used
throughout the paper, the Moyal bracket 
has the expression 
\be
\label{twisted}
(\{g,g^{\prime}\}_M)^{\wedge}(s)=
\frac{2}{\hbar'}\int_{\R^{2n}}\widehat{g}(s^1)
\widehat{g^{\prime}}(s-s^1)
\sin{\left[{\hp}(s-s^1)\wedge s^1/{2}\right]}\,ds^1,
\ee
where, given two vectors $s=(v,w)$ and $s^1=(v^1,w^1)$, 
$s\wedge s^1:=\la w,v_1\ra-\la v,w_1\ra$.\par\noindent
\item   $\{g,g^{\prime}\}_M=\{g,g^{\prime}\}$ if either $g$
or $g^{\prime}$ is quadratic in $(x,\xi)$. 
\end{enumerate}
 The 
 equations (\ref{commu},\ref{Explicitq1},\ref{Explicitq2}) then
become, once written for the symbols:
\begin{eqnarray}
\label{Explicits1}
\sigma(e^{i W(\ep)/\hbar}He^{-i W(\ep)/\hbar})&=&\sum_{l=0}^\infty {\cal H}_l,\; {\cal H}_0:=p_0+\ep
f_0,\; {\cal H}_l:=\frac{\{w,{\cal H}_{l-1}\}_M}{ l}, \;l\geq 1
\\
\Sigma(\ep)&=&\sum_{s=0}^{k}\ep^s p_s +\ep^{k+1}{r}^{(k+1)}
\end{eqnarray}
where
\be
\label{Explicits2}
{p}_s:=\{w_s,p_0\}_M+f_s,\quad s=1,  f_1\equiv f_0
\ee
\begin{eqnarray}
\label{q_s}
f_s &:=&\sum_{r=2}^s\frac{1}{r!}\sum_{{j_1+\ldots+j_r=s}\atop {j_l\geq
1}}\{w_{j_1},\{w_{j_2},\ldots,\{w_{j_r},p_0\}_M\ldots\}_M 
\\
\nonumber
&+&
\sum_{r=2}^{s-1}\frac{1}{r!}\sum_{{j_1+\ldots+j_r=s-1}\atop {j_l\geq
1}}\{w_{j_1},\{w_{j_2},\ldots,\{w_{j_r},f_0\}_M\ldots\}_M, \quad s>1
\end{eqnarray}
In turn, the recursive homological equations become:
\be
\label{qhomsq}
\{w_s,p_0\}_M +f_s=\zeta_s, \qquad \{p_0,\zeta_s\}_M =0
\ee
\par\noindent
{\it 2. Solution of the homological equation and estimates of the solution} 
\vskip 3pt\noindent
$f\in  {\cal A}_{\om,\rho,\sigma}$ clearly entails the existence of the Fourier expansion of $f_{\phi,\om}(u)$, and its uniform
 convergence  with respect 
to $\phi\in\T^2$,  $u$ on compacts of $\R^{2}\times\R^2$, and $\om\in\Gamma$,  namely:
\be
\label{Fourier}
f_{\phi,\om}(u)=\sum_{\nu\in\Z^l}f_{\nu,\om}(u)e^{i\la
\nu,\phi\ra}\Longrightarrow f(u)=\sum_{\nu\in\Z^l}f_{\nu,\om}(u).
\ee
  We further denote, for $\om\in\Gamma$, and $\rho>0$:
\begin{eqnarray}
\|f\|_{\om,\sigma}&:=&\sum_{\nu\in\Z^2}\,\|f_{\nu,\om}\|_{\sigma};\quad {\cal A}_{\om,\sigma}:=\{f(u)\in{\cal
F}_\sigma\,|\,\|f(u)\|_{\om,\sigma} <+\infty
\}
\\
\label{norma2}
\|f\|_{\om,\rho,\sigma}&:=&\sum_{\nu\in\Z^2}\,e^{\rho|\nu|}\|f_{\nu,\om}\|_{\sigma};\quad {\cal A}_{\om,\rho,\sigma}:=\{f(u)\in{\cal
A}_{\om,\sigma}\,|\,\|f(u)\|_{\om,\rho,\sigma} <+\infty
\}
\\\label{norma3}
\|f\|_{\Gamma,\sigma}&:=&\sup_{\om\in\Gamma}\|f\|_{\om,\sigma};\quad {\cal A}_{\Gamma,\sigma}:=\{f(u)\in{\cal
F}_\sigma\,|\,\|f(u)\|_{\Gamma,\sigma} <+\infty
\}
\\
\label{norma5}
\|f\|_{\Gamma,\rho,\sigma}&:=&\sup_{\om\in\Gamma}\|f\|_{\om,\rho,\sigma};
\end{eqnarray}
Hence $ {\cal A}_{\Gamma,\rho,\sigma}=\{f(u)\in{\cal
F}_\sigma\,|\,\|f(u)\|_{\Gamma,\rho,\sigma} <+\infty
\}
$
 and clearly ${\cal A}_{\Gamma,\rho,\sigma}\subset {\cal A}_{\Gamma,\sigma}\subset{\cal F}_\sigma$.   Moreover the following inequalities obviously hold:
\begin{eqnarray}
\label{stima1}
\sup_{u\in\R^2\times \R^2}|f_{\nu,\om}(u)|&\leq& \|\hat{f}_{\nu,\om}(s)\|_{L^1}\leq \|{f}_{\nu,\om}\|_\sigma\leq  \|{f}\|_{\Gamma,\sigma}\leq \|{f}\|_{\Gamma,\rho,\sigma}
\\
\label{stima1bis}
\|\hat{f}\|_{L^1}\leq \|f\|_\sigma &\leq& \|f\|_\sigma \leq \|{f}\|_{\Gamma,\sigma}\leq \|{f}\|_{\Gamma,\rho,\sigma}
\end{eqnarray}

Now the key
remark is that
$\{a,p_0\}_M=\{a,p_0\}$ for any symbol $a$ because $p_0$ is quadratic in $(x,\xi)$. The
homological equation  (\ref{qhomsq}) becomes therefore
\be
\label{chomsq}
\{w_s,p_0\} +f_s=\zeta_s, \qquad \{p_0,\zeta_s\} =0
\ee
We then have:
\par\noindent
\begin{proposition}
\label{lemmaom}
Let $f\in{\mathcal A}_{\Gamma,\rho,\sigma}$. Then the
equation
\be
\label{cchomsq}
\{w,p_0\} +f=\zeta, \qquad \{p_0,\zeta\} =0\ee
admits the solutions $\zeta\in {\mathcal A}_{\Gamma,\sigma}$, $w\in{\mathcal A}_{\Gamma,\rho,\sigma}$
\be
\label{soluzioni}
\label{Zg} \zeta:=f_{0,\om}; \qquad w:=\sum_{\nu\neq
0}\frac{f_{\nu,\om}}{i\la\om,\nu\ra},
\ee 
with the property $\zeta\circ\Psi_{\phi}=\zeta$; i.e., $\zeta$ depends only on
$I_1,I_2$. Moreover:
\begin{eqnarray}
\label{stimaom}
\|\zeta\|_{\Gamma,\sigma}\leq  \|f\|_{\Gamma,\sigma}; \quad 
\|w\|_{\Gamma,\rho,\sigma} \leq \|f\|_{\Gamma,\rho,\sigma}, \quad \|\nabla w\|_{\Gamma,\rho,\sigma} \leq
\frac{4C}{\sigma}\|f\|_{\Gamma,\rho,\sigma}
\end{eqnarray}
for some $C(\Gamma,\delta)>0$.
\end{proposition}
To prove the Proposition we need a preliminary result.
\begin{lemma}
\label{sserie}
Let $w$ be defined by (\ref{soluzioni}), and $\Psi_{\phi,\om}(x,\xi)$  by (\ref{aazione}). Set:
\be
\label{flusso2}
\Xi_{\phi,\om}(x,\xi):=\Phi_{i\phi,i\om}(x,\xi), 
\ee
that is:  $\Xi_{\phi,\om}(x,\xi):=(x^\prime_k,\xi^\prime_k)$, where:
\be
\left\{\begin{array}{l}{\ds x^\prime_k=x_k\cosh\phi_k+\frac{\xi_k}{\om_k}\sinh\phi_k}\\
{\ds \xi^\prime_k=
\xi_k\cosh\phi_k+\om_kx_k\sinh\phi_k}\end{array}\right.\qquad k=1,2
\ee
Then one has, uniformly with respect to $(x,\xi)$ on compacts of $\R^4$:
\begin{eqnarray}
\label{serie1}
w\circ \Psi_{\phi,\om}(x,\xi) &=&  \sum_{\nu\neq
0}\frac{f_{\nu,\om}(x,\xi)}{i\la\om,\nu\ra}e^{i\la\nu,\phi\ra},\quad \phi\in\T^2
\\
\label{serie2}
w\circ \Xi_{\phi,\om}(x,\xi) &=&  \sum_{\nu\neq
0}\frac{f_{\nu,i\om}(x,\xi)}{\la\om,\nu\ra}e^{-\la\nu,\phi\ra}, \quad |\phi|\leq \rho-\eta, \;\forall\,0<\eta<\rho
\end{eqnarray}
Moreover  there is $C(\delta)>0$ such that:
\be
\label{stimaw}
\|w\|_{\om,\rho,\sigma}\leq C\|f\|_{\om,\rho,\sigma}; \quad \|w\|_{i\om,\rho,\sigma}\leq C\|f\|_{i\om,\rho,\sigma}
\ee
\end{lemma}
{\bf Proof}   
\newline
Let us first prove that (\ref{soluzioni}), whose convergence is proved below,  solves
(\ref{cchomsq}), and that $w\circ \Psi_{\phi,\om}(x,\xi)$ admits the representation (\ref{serie1}).  Following the argument of (\cite{BGP}), Lemma 3.6,  let us write:
\begin{eqnarray*}
\{p_0,pw\}(x,\xi)=\left.\frac{d}{dt}\right|_{t=0}w\circ\Psi_{\om t,\om}(x,\xi)=\left.\frac{d}{dt}\right|_{t=0} \sum_{0\neq\nu\in\Z^2}\,\frac{f_{\nu,\om}\circ \Psi_{\om t,\om}(u)}{i\la\om,\nu\ra}
\\
=\left.\frac{d}{dt}\right|_{t=0} \sum_{0\neq\nu\in\Z^2}\,\frac{f_{\nu,\om}\circ \Psi_{\om t,\om}(u)}{i\la\om,\nu\ra}=\left.\frac{d}{dt}\right|_{t=0} \sum_{0\neq\nu\in\Z^2}\,\frac{f_{\nu,\om}(u)e^{i\la \nu,\om t\ra}}{i\la\om,\nu\ra}=
\sum_{0\neq\nu\in\Z^2}\,f_{\nu,\om}(u)\end{eqnarray*}
Clearly, this equality also  entails $\zeta=f_{0,\om}$.  Consider now the expansions (\ref{serie1}, \ref{serie2}).  First, it is easy to check that $\om\in\Gamma$ if and only if  $i\om\in\Gamma$. Now we have:
$$
w_{\nu,\om}=\frac{f_{\nu,\om}(x,\xi)}{i\la\om,\nu\ra}
$$
and therefore, by a straightforward application of Lemma \ref{larged}:
$$
\|w_{\nu,\om}\|_\sigma \leq C\|f_{\nu,\om}\|_\sigma .
$$
Hence:
$$
\|w\|_{\omega,\rho,\sigma}=\sum_{\nu\in\Z^2}\,e^{\rho |\nu|}\|w_{\nu,\om}\|_\sigma\leq C \sum_{\nu\in\Z^2}\,e^{\rho |\nu|}\|f_{\nu,\om}\|_\sigma =\|f\|_{\omega,\rho,\sigma}\quad\forall\,\om\in\Gamma.
$$
Therefore $q\in {\mathcal A}_{\Gamma,\rho,\sigma}$ entails $w\circ\Psi_{\om,\phi}\in {\mathcal A}_{\Gamma,\rho,\sigma}$, whence the uniform convergence of the series (\ref{serie1}).  Now $i\om\in\Gamma$ if $\om\in\Gamma$; hence  $w\circ\Psi_{i\om,\phi}\in {\mathcal A}_{\Gamma,\rho,\sigma}$. On the other hand, the replacement  $\phi\to i\phi$ maps $\Psi_{\phi,i\om}(x,\xi)$ into $\Xi_{\phi,\om}(x,\xi)$, and the series (\ref{serie2}) is uniformly convergent  if $|{\rm Im}\,\phi |<\rho-\eta$, $0<\eta<\rho$. Formula (\ref{serie2}) is therefore proved. 
This concludes the proof of  the Lemma.
\vskip 0.3cm\noindent
{\bf Proof of Proposition 2.1}  
\newline
Let us first prove that  $\zeta$ depends only on $I_1,I_2$. Consider for the sake of simplicity $u=(x,\xi)\in\R^2$. Since $f\in{\mathcal A}_{\Gamma,\rho,\sigma}$, we can write:
\begin{eqnarray*}
f_{\phi,\om}(x,\xi)=
\sum_{m.n=0}^\infty\,\frac{a_{mn}}{2^{m+n}}\left[(x+\frac{\xi}{i\om})e^{i\phi}+
(x-\frac{\xi}{i\om})e^{-i\phi}\right]^m\left[(-i\om x+\xi)e^{i\phi}+
(i\om x +\xi)e^{-i\phi}\right]^n
\end{eqnarray*}
The average over $\phi$ eliminates all terms but those  proportional to 
$$
[(x+\frac{\xi}{i\om})(x-\frac{\xi}{i\om})]^k[(-i\om x+\xi)(i\om x +\xi)]^l
$$ 
i.e. to  $I^kI^l$. The estimate
$\|\zeta\|_{\om,\sigma}\leq 
\|f\|_{\om,\sigma}$ is obvious, and entails $\|\zeta\|_{\Gamma,\sigma}\leq 
\|f\|_{\Gamma,\sigma}$. The second estimate in (\ref{stimaom}) has been proved in Lemma 2.1 above. To prove the third one,   consider the function $f\circ\Psi_{\phi,\om}(z)$ and compute,
for $j=1,2$:
\begin{eqnarray*}
\frac{d}{d\phi_j}w\circ\Psi
_{\phi,\om}(z)|_{\phi=0}&=&\left.\frac{\partial w}{\partial x_j}
\frac{\partial x^\prime_j}{\partial \phi_j}+\frac{\partial w}{\partial
\xi_j}
\frac{\partial \xi^\prime_j}{\partial \phi_j}\right|_{\phi=0}
=
\\
\frac{\partial w}{\partial
x_j}\frac{\xi_j}{\om_j}-
\frac{\partial w}{\partial \xi_j}\om_jx_j&=&\sum_{0\neq \nu\in\Z^2}
\frac{\nu_jf_{\nu,\om}}{i\la\om,\nu\ra}
\end{eqnarray*}
Therefore, once more by Lemma \ref{larged}, 
\begin{eqnarray*}
\left\|\frac{\partial w}{\partial
x_j}\frac{\xi_j}{\om_j}-
\frac{\partial w}{\partial \xi_j}\om_jx_j\right\|_{\om,\rho,\sigma} &\leq& \sum_{0\neq \nu\in\Z^2}\,e^{\rho|\nu|}
\frac{|\nu_j|}{|\la\om,\nu\ra|}
\|f_{\nu,\om}\|_{\om,\sigma}
\\
&\leq& C \sum_{0\neq \nu\in\Z^2}\,e^{\rho|\nu|}
\|f_{\nu,\om}\|_{\om,\sigma}= C\|f\|_{\om,\rho,\sigma}. 
\end{eqnarray*}
This yields:
\be
\label{stima5}
\left\|\frac{\partial w}{\partial
x_j}\frac{\xi_j}{\om_j}-
\frac{\partial w}{\partial \xi_j}\om_jx_j\right\|_{\Gamma,\rho,\sigma} \leq 
C\|f\|_{\Gamma,\rho,\sigma}.
\ee
In the same way:
\begin{eqnarray*}
\frac{d}{d\phi_j}w\circ\Xi
_{\phi,\om}(z)|_{\phi=0}&=&\left.\frac{\partial w}{\partial x_j}
\frac{\partial x^\prime_j}{\partial \phi_j}+\frac{\partial w}{\partial
\xi_j}
\frac{\partial \xi^\prime_j}{\partial \phi_j}\right|_{\phi=0}
=
\\
\frac{\partial w}{\partial
x_j}\frac{\xi_j}{\om_j}+
\frac{\partial w}{\partial \xi_j}\om_jx_j&=&\sum_{0\neq \nu\in\Z^2}
\frac{\nu_jf_{\nu,i\om}}{\la\om,\nu\ra}
\end{eqnarray*}
whence, by Lemma \ref{larged}, 
\begin{eqnarray*}
\left\|\frac{\partial w}{\partial
x_j}\frac{\xi_j}{\om_j}+
\frac{\partial w}{\partial \xi_j}\om_jx_j\right\|_{i\om,\rho,\sigma} &\leq& \sum_{0\neq \nu\in\Z^2}\,e^{\rho|\nu|}
\frac{|\nu_j|}{|\la\om,\nu\ra|}
\|f_{\nu,i\om}\|_{i\om,\sigma}
\\
&\leq& C \sum_{0\neq \nu\in\Z^2}\,e^{\rho|\nu|}
\|f_{\nu,i\om}\|_{i\om,\sigma}= C\|f\|_{i\om,\rho,\sigma}. 
\end{eqnarray*}
Recalling that $\om\in\Gamma$ if and only if $i\om\in\Gamma$ we get:
\be
\label{stima6}
\left\|\frac{\partial w}{\partial
x_j}\frac{\xi_j}{\om_j}+
\frac{\partial w}{\partial \xi_j}\om_jx_j\right\|_{\Gamma,\rho,\sigma} \leq 
C\|f\|_{\Gamma,\rho,\sigma}.
\ee
Denote now $s_j$, $t_j$ the Fourier dual variables of
$(x_j,\xi_j)$,
$j=1,2$.  Then, by definition (we drop for the sake of simplicity the dependence of $\om$):
$$
\left\|\frac{\partial w}{\partial
x_j}{\xi_j}\right\|_{\sigma}=\int_{\R^4}\left|s_j\frac{\partial
\widehat{w}(s_j,t_j)}{\partial t_j}\right|e^{\sigma(|s|+|t|)}\,dsdt
$$
Applying Lemma \ref{Poinc} to the integration over $t_j$ we get:
\begin{eqnarray*}
\left\|\frac{\partial w}{\partial
x_j}\right\|_{\om,\sigma}&=&\sum_{\nu\in\Z^2}\,\int_{\R^4}\left|s_j
\widehat{w}_{\nu,\om}(s_j,t_j)\right |e^{\sigma(|s|+|t|)}\,dsdt \leq 
\\
&\leq& \frac{2}{\sigma}\sum_{\nu\in\Z^2}\,\int_{\R^4}\left|s_j\frac{\partial
\widehat{w}_{\nu,\om}(s_j,t_j)}{\partial t_j}\right|e^{\sigma(|s|+|t|)}\,dsdt
\\
&=&
\frac{2}{\sigma}\sum_{\nu\in\Z^2}\,\left\|\frac{\partial w_{\nu,\om}}{\partial
x_j}{\xi_j}\right\|_\sigma
= \frac{2}{\sigma}\left\|\frac{\partial w}{\partial
x_j}\xi_j\right\|_{\om,\sigma}
\end{eqnarray*}
Therefore, by (\ref{stima5},\ref{stima6})
$$
\left\|\frac{\partial w}{\partial
x_j}\right\|_{\Gamma,\rho,\sigma}\leq \frac{2C|\om_j|}{\sigma}\|f\|_{\Gamma,\omega,\sigma}
$$
Analogously, applying this time Lemma \ref{Poinc} to the integration over $s_j$:
$$
\left\|\frac{\partial w}{\partial \xi_j}\right\|_{\Gamma,\rho,\sigma}\leq
\frac{2C}{\sigma{|\om_j|}}\|f\|_{\Gamma,\omega,\sigma}.
$$
This is enough to prove the Proposition. 
\vskip 0.3cm\noindent
{\it  3. Iterative Lemma}
\begin{proposition}
Set:
$$ 
\mu:=\frac{4\ep\|{f}_0\|_{\Gamma,\rho,\sigma}}{\sigma}. 
$$
Let $\mu<1/4$ and consider for
$k=1,2,\ldots$ the function
\begin{eqnarray}
\label{sigmak}
\Sigma_k:=p_0+\ep{\cal Z}_k+v_k
\end{eqnarray}
with ${\cal Z}_k, v_k\in{\mathcal A}_{\Gamma,\rho,\sigma}$, and let ${\cal Z}_k$ depend on $(I_1,I_2)$ only. 
\newline
Assume moreover:
\begin{eqnarray}
\label{iter1}
\|{\cal Z}_k\|_{\Gamma,\sigma}&\leq& \left\{\begin{array}{cc} \ds 0 & {\rm if}\;k=0
\\
\ds \sum_{s=0}^{k-1}(2\mu)^s &{\rm if}\;k\geq 1
\end{array}\right.
\\
\label{iter2}
\|v_k\|_{\Gamma,\rho,\sigma} &\leq&\ep(2\mu)^k\|f_0\|_{\Gamma,\rho,\sigma}
\end{eqnarray}
Let $S_k$ be the Weyl quantization of $\Sigma_k$. Then there exists a unitary map
$T_k:L^2\to L^2$, $\ds T_k:=e^{i\ep W/\hbar}$ such that the Weyl symbol of the
transformed operator
$T_kS_kT^\ast_k:=S_{k+1}$ is given by (\ref{sigmak}) with $k+1$ in place of $k$ and satisfies 
(\ref{iter1},\ref{iter2}) with $k+1$ in place of $k$. 
\end{proposition}
{\bf Proof} As in \cite{BGP}, Proposition 3.2, the
homological equation:
\be
\label{hom5}
\{p_0,w\}+v_k=\ep {\cal V}_k
\ee
determines the symbol $w$ of $W$. Here the second unknown ${\cal V}_k$ has to depend
on
$(x,\xi)$ only through $I_1,I_2$. 
Applying 
Proposition 1 
we find that 
$w$ and
${\cal V}_k$ exist and fulfill the estimates
$$
\|{w}\|_{\Gamma,\rho,\sigma}\leq \ep \|{f}_0\|_{\Gamma,\rho,\sigma}(2\mu)^k ;\quad \|{\nabla w}\|_{\Gamma,\rho,\sigma}\leq
(2\mu)^{k+1};\quad\|{\cal V}_k\|_{\Gamma,\rho,\sigma}\leq \|{f}_0\|_{\Gamma,\rho,\sigma}(2\mu)^k.
$$
Define now:
\begin{eqnarray*}
{\cal Z}_{k+1}&:=&{\cal Z}_{k}+{\cal V}_{k};\quad v_{k+1}:=\ep\sum_{l\geq 1}{\cal Z}_{k}^l+
\sum_{l\geq 1}{v}_{k}^l+\sum_{l\geq 1}p_{l0}
\\
{\cal Z}_{k}^0&:=&{\cal Z}_{k};\quad {\cal Z}_{k}^l:=\frac1{l}\{w,{\cal Z}_{k}^{l-1}\}_M
\end{eqnarray*}
and analogous definitions for $v_k^l$ and $p_{l0}$.  Clearly $v_{k+1}\in {\cal A}_{\Gamma,\rho,\sigma}$ by Lemma \ref{lemmaiter} 
below. Then the symbol of the transformed operator has the form (\ref{sigmak}) with $k+1$ in
place of
$k$.  To get the estimates, for
$k\geq 1$  we can write, by Proposition 1 and Lemmas 2.2. 2.3. 2.4:
\begin{eqnarray*}
\sum_{l\geq 1}\|({v}_{k}^l)\|_{\Gamma,\rho,\sigma}&\leq& \ep(2\mu)^k\sum_{l\geq
1}(2\mu)^l=\frac{\ep(2\mu)^{k+1}}{1-2\mu}\leq \ep (2\mu)^{k+1}
\\
\sum_{l\geq 1}\|{{\cal Z}_{k}^l}\|_{\Gamma,\sigma}&\leq &\|{{\cal
Z}_{k}^l}\|_{\Gamma,\sigma}\cdot\frac{\mu}{1-\mu}\leq 2\mu, \quad \sum_{l\geq
2}\|{{p}_{l0}}\|_{\Gamma,\rho,\sigma}\leq \ep(2\mu)^{k+1}
\end{eqnarray*}
whence the assertion in a straightforward way. 
\par\noindent
{\bf Proof of Theorem 1} 
\par\noindent
By Proposition 2 there is $\ep^\ast>0$ such that 
$$
\lim_{k\to\infty} p_0+\ep {\cal Z}_k:=\Sigma(\ep)
$$
exists in the  $\\|\cdot\|_{\Gamma,\rho,\sigma}$ norm if $|\ep|<\ep^\ast$. Then $S(\ep):=Op_h^W(\Sigma(\ep))$ is
unitarily equivalent to $H(\ep)$. Since ${\cal Z}_k$ is a polynomial of order $k-1$ in
$\ep$, we can write $\ds \Sigma_k=p_0+\sum_{l=1}^k\zeta^{(l)}\ep^l+v_k$, where
$\zeta^{(l)}(I_1,I_2)$ are solutions of the homological equations (\ref{qhomsq}); therefore
$S(\ep)$  has the form (\ref{passo1bis}). Note that $\ds
\lim_{k\to\infty}\|v_k\|_{\Gamma,\rho,\sigma}=0$ entails $\ds \lim_{k\to\infty}\|R_k\|_{L^2\to
L^2}=0$. To sum up, the Weyl symbol
$\Sigma(\ep,\hbar)$ has the  convergent (uniform with respect to $\hbar$)  normal form 
$$
\Sigma(\ep,\hbar)=p_0(I)+\sum_{n=1}^\infty {\cal Z}_n(I,\hbar)\ep^n
$$
Then the assertions of Theorem 1 follow exactly as in \cite{Sj} (see also \cite{BGP}). This
concludes the proof. 
\vskip 0.3cm\noindent
{\it  4. Auxiliary results}
\begin{lemma} 
\label{Mo1}
Let $(g,g^{\prime}, \nabla g,\nabla g^{\prime})\in {\cal
F}_{\sigma}$. Then:
\be
\|\{g,g^{\prime}\}_M\|_{\sigma} \leq
\|\nabla g\|_{\sigma}
\|\nabla g^{\prime}\|_{\sigma}.
\ee
If $(g,g^{\prime}, \nabla g,\nabla g^{\prime})\in {\cal
A}_{\om,\rho,\sigma}$ then
\be
\|\{g,g^{\prime}\}_M\|_{\om,\rho,\sigma} \leq
\|\nabla g\|_{\om,\rho,\sigma}
\|\nabla g^{\prime}\|_{\om,\rho,\sigma}.
\ee
and if  $(g,g^{\prime}, \nabla g,\nabla g^{\prime})\in {\cal
A}_{\Gamma,\rho,\sigma}$:
\be
\|\{g,g^{\prime}\}_M\|_{\Gamma,\rho,\sigma} \leq
\|\nabla g\|_{\Gamma,\rho,\sigma}
\|\nabla g^{\prime}\|_{\Gamma,\rho,\sigma}.
\ee
\end{lemma}
{\bf Proof} 
\newline
We repeat the argument of \cite{BGP}, Lemma 3.1. We have $\ds |s\wedge s^1|\leq |s|\cdot |s^1|$. Hence by
(\ref{twisted}) and the definition of the $\sigma-$ norm we get:
\begin{eqnarray*}
\|\{g,g^{\prime}\}_M\|_{\sigma} &=&
\frac{2}{\hbar}\int_{\R^{2l}}e^{\sigma|s|}\,ds\int_{\R^{2l}}|\hat g
(s)\hat{g^{\prime}}(s-s^1)|\cdot |{\rm sinh}(\hbar (s-s^1)\wedge
s^1)/2|\,ds^1  
\\
&\leq&
\frac{2}{\hbar}\int_{\R^{2l}}\,ds\int_{\R^{2l}}e^{\sigma(|s|+|s^1|)}|\hat g
(s)\hat{g^{\prime}}(s^1)|\cdot |{\rm sinh}(\hbar s\wedge s^1)/2|\,ds^1
\\
&\leq&
\int_{\R^{2l}}e^{\sigma|s|}|\hat
g (s)|\,ds
\int_{\R^{2l}}e^{\sigma|s^1|}|\hat{g^{\prime}}(s^1)|\cdot |
s\wedge s^1|\,ds^1=
\\
&\leq&
\int_{\R^{2l}}e^{\sigma|s|}|\hat
g (s)||s|\,ds
\int_{\R^{2l}}e^{\sigma|s^1|}|\hat{g^{\prime}}(s^1)|\cdot | s^1|\,ds^1
=\|\nabla g\|_{\sigma}
\|\nabla g^{\prime}\|_{\sigma}.
\end{eqnarray*}
The remaining two inequalities follow from the first one by exactly the same argument of \cite{BGP}, Lemma 3.4. This concludes the proof of the Lemma. 
\vskip 0.2cm\noindent
\begin{lemma} 
\label{Poinc}
Let $g\in {\cal
F}_{\sigma}$, $u=(x,\xi)\in\R^{2l}$. Then:
\be
\label{Po}
\|g\|_\sigma\leq \frac1{\sigma}\|ug\|_\sigma
\ee
\end{lemma}
{\bf Proof} 

\noindent
Setting $f(s):=\hat{g}(s)$ (\ref{Po}) is clearly 
equivalent to 
\be
\label{Po1}
\int_{\R^{2l}}e^{\sigma|s|}|
f(s)|\,ds\leq 
\frac1{\sigma}\int_{\R^{2l}}e^{\sigma|s|}|
\nabla f(s)|\,ds
\ee
We may limit ourselves to prove this inequality in the one-dimensional case, namely to
show that:
\be
\label{Po2}
\int_{\R}e^{\sigma|s|}|
f(s)|\,ds\leq 
\frac1{\sigma}\int_{\R}e^{\sigma|s|}|f^{\prime}(s)|\,ds
\ee
To see this, first write, for $s>0$:
$$
e^{\sigma s}
f(s)=-\int_s^\infty e^{\sigma t}f^\prime(t)e^{\sigma(s-t)}\,dt
$$
whence, for $A>0$:
\begin{eqnarray*}
\int_A^\infty |e^{\sigma s}
f(s)|\,ds &\leq& \int\!\!\!\int_{A\leq s\leq t\leq\infty}\,|f^\prime(t)|e^{\sigma
s}\,dsdt=\int_A^\infty|f^\prime(t)|\int_A^t e^{\sigma
s}\,dsdt=
\\
&=&\sigma^{-1}\int_A^\infty|f^\prime(t)|(e^{\sigma t}-e^{\sigma A})\,dt\leq 
\sigma^{-1}\int_A^\infty|f^\prime(t)|e^{\sigma t}\,dt
\end{eqnarray*}
Likewise, for $s<0$, $A<0$:
$$
e^{-\sigma s}
f(s)=\int_{-\infty}^s e^{-\sigma t}f^\prime(t)e^{-\sigma(s-t)}\,dt
$$
\begin{eqnarray*}
\int_{-\infty}^A |e^{-\sigma s}
f(s)|\,ds &=& \int\!\!\!\int_{-\infty\leq t\leq s\leq A}\,|f^\prime(t)|e^{-\sigma
s}\,dsdt=\int_{-\infty}^A|f^\prime(t)|\int_t^A e^{-\sigma
s}\,dsdt=
\\
&=&\sigma^{-1}\int_{-\infty}^A|f^\prime(t)|(e^{-\sigma t}-e^{-\sigma A})\,dt \leq 
\sigma^{-1}\int_{-\infty}^A|f^\prime(t)|e^{-\sigma t}\,dt
\end{eqnarray*}
Performing the limit $A\to 0$ in both inequalities we get (\ref{Po2}). This concludes the proof of the Lemma.
\vskip 0.2cm\noindent
\begin{lemma}
\label{lemmaiter}
\par\noindent
 Let $g\in {\mathcal A}_{\Gamma,\rho,\sigma}$,  $w\in {\mathcal A}_{\Gamma,\rho,\sigma}$.
\par\noindent
1.  Define
$$ 
g_r:=\frac{1}{r}\{w,g_{r-1}\}_{M}, \qquad r\geq 1; \;\;g_0:=g.
$$ 
Then  $g_r\in  {\mathcal A}_{\Gamma,\rho,\sigma}$ 
and
the following estimate holds
\be
\label{stimaindiv}
\|{g_r}\|_{\Gamma,\rho,\sigma}\leq \left(4\frac{{\nabla w}_{\Gamma,\rho,\sigma}}{\sigma}\right)^r
\|{g}\|_{\Gamma,\rho,\sigma}.
\ee
2. Let   $w$ solve the
homological equation (\ref{qhomsq}). Define the sequence $p_{r0}:
r=0,1,\ldots$:
$$ 
p_{00}:=p_0; \qquad p_{r0}:=\frac{1}{r}\{w,p_{r-10}\}_M, \;r\geq 1.
$$ 
Then  $p_{r0}\in\As$ and fulfills the following estimate
\be
\label{stimaindiv1}
\|{p_{r0}}\|_{\Gamma,\rho,\sigma}\leq
\left(4\sigma^{-1}\|{\nabla w}\|_{\Gamma,\rho,\sigma}\right)^{r-1}\|{f_0}\|_{\Gamma,\rho,\sigma},
\quad r\geq 1.
\ee
\end{lemma} 
{\bf Proof} 
\par\noindent
Both estimates (\ref{stimaindiv},\ref{stimaindiv1}) are straightforward consequences of Lemmas 2.2 and 2.3:  as far as (\ref{stimaindiv1}) is concerned, it is indeed enough to note that $\{w,p_0\}=\zeta-q$ whence 
$$
\|{p_{10}}\|_{\Gamma,\rho,\sigma}+\|{\nabla p_{10}}\|_{\Gamma,\rho,\sigma}\leq \frac{4\|{f_0}\|_{\Gamma,\rho,\sigma}}{\sigma}.
$$
\vskip 0.2cm\noindent
\begin{lemma}
\label{larged}
If (A3) holds there is $C_\delta>0$ independent of $\om\in\Gamma$ such that
\be
\label{grandid} |\om_1\nu_1+\om_2\nu_2|\geq C_\delta\sqrt{v_1^2+\nu_2^2}
\ee
\end{lemma}
{\bf Proof} \newline
We have to show the existence of $C_\delta>0$ such that
\be
\label{min}
f(\nu_1,\nu_2):=\frac{|\om_1\nu_1+\om_2\nu_2|^2}{v_1^2+\nu_2^2}\geq
C_\delta,\quad \forall\,(\nu_1,\nu_2)\in\Z^2, (\nu_1,\nu_2)\neq (0,0)
\ee
Notice that $f$ is homogeneous of degree $0$, namely
$f(\mu\nu_1,\mu\nu_2)=f(\nu_1,\nu_2)$ $\forall\,(\nu_1,\nu_2)\in\Z^2,
(\nu_1,\nu_2)\neq (0,0)$, $\forall\,\mu\in\R$, $\mu\neq 0$. Hence it is
enough to show that
\be
\label{min1}
F(x,y):={|\om_1x+\om_2 y|^2}\geq
C_\delta,\quad \forall\,(x,y)\in S^1
\ee
or, writing $x=\cos\theta, y=\sin\theta$:
$$
F(\theta):=\frac12[|\om_1|^2+|\om_2|^2]+\frac12[|\om_1|^2-|\om_2|^2]\cos{2\theta}
+\la\om_1,\om_2\ra\sin{2\theta}\geq C
$$
Note that $F(0)=F(2\pi)=|\om_1|^2$.  A simple study of the function
$F(\theta): S^1\to\R$ under the assumption (A2) shows the existence of
$C_\delta\downarrow 0$ as $\delta\uparrow 1$ such that $|F(\theta)|\geq
C_\delta$ $\forall\,\theta\in S^1$. We omit the elementary details.
\vskip 1.0cm\noindent
{\bf \Large Appendix}  
\par\noindent
Consider the function  $f:\C^4\to \R$
$$
f(z):=e^{-|z|^2}P_n(z), \quad z\in\C^4,\;|z|=\sum |z_k|^2.
$$
Here $P_n(z)$ is a polynomial of degree $n$.
\par\noindent
Let us verify that $f$ belongs to ${\cal A}_{\Gamma,\rho,\sigma}$; namely, there are $\rho>0$, $\sigma>0$ such that:
$$
\sup_{\om\in\Gamma}\sum_{\nu\in\Z^2}\,e^{\rho |\nu|}\|f_{\nu,\om}(u)\|_\sigma <+\infty.
$$
 It is  clearly enough to consider the case $u=(x,\xi)\in\R^2$, $n=0$.  
 \newline
 Set: $\om:=\gamma e^{i\theta}, 0\leq \theta\leq 2\pi$, $\delta_1\leq \gamma\leq \delta_2$. Then:
 $$
 |\Psi_{\phi,\om}(u)|^2=\left|x{\cos}\phi+\frac{\xi}{\om}{\sin}\phi\right|^2+|\xi{\cos}\phi-\om x{\sin}\phi|^2=Ax^2+Bx\xi+C\xi^2
 $$
  $$
  A:=\cos^2\phi+\gamma^2\sin^2\phi; \quad B:=\cos\theta(\gamma^{-1}-\gamma)\sin2\phi, \quad C:=\cos^2\phi+\gamma^{-2}\sin^2\phi
  $$
  Therefore we can write:
\begin{eqnarray*}
f_{\phi,\om}(u)&:=&f\circ \Psi_{\om,\phi}(u)= e^{-\la Q(\gamma,\theta,\phi)u,u\ra},
\qquad
 Q(\gamma,\theta,\phi):=\left(\begin{array}{cc} \ds A & \frac12 B \\ \frac12 B & C\end{array}\right)
 \\
 \det{Q}&=&\cos^4\phi+\sin^4\phi+[(\gamma^{-2}+\gamma^2)-\cos^2\theta(\gamma^{-1}-\gamma)^2]\sin^2\phi\cos^2\phi
 \\
&=& 1+\kappa (1-{\cos}^2\theta){\sin}^2\phi{\cos^2}\phi 
\\
 {\rm Tr}\,{Q}&=& 2+\kappa{\sin}^2\phi
 \\
\kappa&:=&\gamma^{-2}+\gamma^2-2\geq 0 
\end{eqnarray*}
whence, $\forall\,(\theta,\phi)\in [0,2\pi]\times [0,2\pi]$ 
\begin{eqnarray*}
1\leq \lambda_1\lambda_2 \leq 1+\kappa, 
\quad
2\leq \lambda_1+\lambda_2\leq 2+\kappa
\end{eqnarray*}
where $0<\lambda_{1}(\gamma,\theta,\phi)\leq \lambda_{2}(\gamma,\theta,\phi)$ denote the eigenvalues of $Q(\gamma,\theta,\phi)>0$. This easily yields the uniform estimate: 
$$
\frac1{D}\leq \lambda_{1}(\gamma,\theta,\phi)\leq \lambda_{2}(\gamma,\theta,\phi)\leq  D, \quad D:=\frac12[2+\kappa+\sqrt{(2+\kappa)^2-4}].
$$
 \newline 
Consider now the Fourier coefficients $f_{\nu,\om}(u)=f_{\nu,\gamma,\theta}(u)$:
\begin{eqnarray*}
f_{\nu,\gamma,\theta}(u):=\frac1{2\pi}\int_0^{2\pi}f\circ \Psi_{\om,\phi}(u)e^{-i\nu\phi}\,d\phi=
\frac1{2\pi}\int_0^{2\pi}e^{-\la Q(\gamma,\theta,\phi)u,u\ra}e^{-i\nu\phi}\,d\phi
\end{eqnarray*}
and compute  their Fourier transform:
\begin{eqnarray*}
\hat{f}_{\nu,\gamma,\theta}(s)
&=&\frac1{2(\pi)^2}\int_{\R^2}\int_0^{2\pi}e^{-\la Q(\gamma,\theta,\phi)u,u\ra}e^{-i\nu\phi}e^{-i\la u,s\ra}\,d\phi\,du 
\\
&=&\frac{2}{(2\pi)^2 \sqrt{\det{Q}} }\int_0^{2\pi}e^{-\la Q^{-1}(\gamma,\theta,\phi)s,s\ra/2}e^{-i\nu\phi}\,d\phi, \quad s\in\R^2
\\
Q^{-1}(\gamma,\theta,\phi)&=&\frac{1}{\det{Q}}\left(\begin{array}{cc} \ds C & -\frac12 B \\ -\frac12 B & A\end{array}\right).
\end{eqnarray*}
Since  
$$
\la s,Q^{-1}(\gamma,\theta,\phi)s\ra\geq \lambda_2^{-1}s^2\geq\frac{s^2}{D}
$$ 
$\forall\,(\theta,\phi)\in[0,2\pi]\times [0,2\pi]$ 
 we get  the $(\nu,\theta,\phi)$-independent estimate
$$
|\hat{f}_{\nu,\gamma,\theta}(s)|\leq \frac{2}{(2\pi)^2}e^{-|s|^2/D}\int_0^{2\pi}\,d\phi=\frac{1}{\pi}e^{-|s|^2/D}
$$
Therefore $\|f_{\nu,\om}\|_\sigma<+\infty$ $\forall\,\sigma >0$, $\forall\,\nu\in\Z^2$. 
\par\noindent
Let now $\phi\in\C$.  Writing:
$$
\det{Q(\gamma,\theta,\phi)}=1+\frac{A(\gamma,\theta)}{4}\sin^2(2\phi),\quad A(\gamma,\theta):=\kappa(1-\cos^2\theta)\geq 0
$$
we get (omitting the elementary details):
$$
\det{Q(\gamma,\theta,\phi)}\neq 0,\qquad |{\rm Im}\,\phi |<\frac{1}{4}{\rm arc cosh}(1+8/\kappa)\,.
$$
Therefore the function
$$
\phi\mapsto \frac{e^{-\la Q^{-1}(\gamma,\theta,\phi)s,s\ra}}{\sqrt{\det{Q(\gamma,\theta,\phi)}}}:=G_{\gamma,\theta,s}(\phi)
$$
is analytic with respect to $\phi$ in the strip $\ds |{\rm Im}\,\phi |<\frac{1}{4}{\rm arc cosh}(1+8/\kappa):=m(\kappa)$ uniformly with respect to $(\gamma,\theta,s)\in [\delta_1,\delta_2]\times [0,2\pi]\times \R^2$.  \newline 
 In  turn the analyticity entails, as is well known, that for any $\ds 0<\eta< m(\kappa)$ there exists    $\ds \rho_1>m(\kappa)-\eta $ independent of $(\gamma,\theta,s)\in [\delta_1,\delta_2]\times [0,2\pi]\times \R^2$ such that 
$$
|\hat{f}_{\nu,\gamma,\theta}(s)| \leq \sup_{|{\rm Im}\,\phi |\leq \eta}|G_{\gamma,\theta,s}(\phi)|{e^{-\rho_1|\nu|}}.
$$
Since $\ds \det{Q(\gamma,\theta,\phi)}\neq 0$ for $ |{\rm Im}\,\phi |\leq \eta$, 
  there exist $K_1(\eta)>, K_2(\eta)>0$ independent of $(\gamma,\theta)$ such that:
  $$
| \la Q^{-1}(\gamma,\theta,\phi)s,s\ra |\geq K_1|s|^2, \quad \frac{1}{|\sqrt{\det{Q(\gamma,\theta,\phi)|}}}<K_2(\eta)
$$
and therefore
$$
|\hat{f}_{\nu,\gamma,\theta}(s)|\leq\frac{K_2(\eta)}{2\pi}e^{- K_1|s|^2}{e^{-\rho_1|\nu|}}.
$$
This in turn entails the existence of $K_3(\eta)>0$ independent of $\nu$ such that, $\forall\,\sigma >0$:
$$
\|f_{\nu,\om}\|_\sigma =\int_{\R^2}e^{\sigma |s|} |\hat{f}_{\nu,\gamma,\theta}(s)|\,ds \leq K_3 e^{-\rho_1|\nu|}.
$$
Hence, $\forall\,0<\rho<\rho_1$:
$$
\|f\|_{\om,\rho,\sigma}=\sum_{\nu\in\Z^2}e^{\rho|\nu|}\|f_{\nu,\om}\|_\sigma<K(\eta)
$$
for some $K(\eta)>0$ independent of $\omega\in\Gamma$.  We can thus conclude that
$$
\|f\|_{\Gamma,\rho,\sigma}=\sup_{\omega\in\Gamma}\sum_{\nu\in\Z^2}e^{\rho|\nu|}\|f_{\nu,\om}\|_\sigma<K
$$
i.e.,  $f\in{\cal A}_{\Gamma,\rho,\sigma}$. 
\par\noindent
{\bf Remark}
\newline
We have checked that $f\in {\cal A}_{\Gamma,\rho,\sigma}$.  This entails 
$f\in{\cal F}_{\sigma}$. By the Paley-Wiener theorem, $\ds f_{\phi,\om}(u)=e^{-(Ax^2+Bx\xi+C\xi^2)}$ must have, $\forall\,(\phi,\om)$, a holomorphic continuation $ g_{\phi,\om}(z_1,z_2)$ from  $u=(x,\xi)\in\R\times\R$ to $z=(z_1,z_2)=(x+iy,\xi+i\eta)\in\C\times \C$. This holomorphic continuation is clearly
$$
g_{\phi,\om}(z_1,z_2):=
e^{-Az_1^2+Bz_1z_2+Cz_2^2}.
$$
$g_{\phi,\om}(z_1,z_2)$ of course does not coincide with 
$$
f\circ\Psi_{\phi,\om}((z_1,z_2))=\exp{\{-[|z_1{\cos}\phi+\frac{z_2}{\om}{\sin}\phi |^2+|z_2{\cos}\phi-\om z_1{\sin}\phi |^2 ]\}}
$$ 
when $(y,\eta)\neq (0,0)$. 
\vfill

\end{document}